\documentstyle[epsf]{aipproc}

\newcommand{\rll}{\langle}
\newcommand{\rl}{\rangle}

\begin{document}
\title{Level Densities by Particle-Number Reprojection Monte Carlo Methods}

\author{S. Liu$^1$, Y. Alhassid$^1$ and H. Nakada$^2$}
\address{$^1$Center for Theoretical Physics, Sloane Physics Laboratory,
Yale University, New Haven, Connecticut 06520 \\
$^2$Department of Physics, Chiba University,
Inage, Chiba 263-8522, Japan}

\maketitle

\begin{abstract}
A particle-number reprojection method is applied in the framework of
the shell model Monte Carlo approach to calculate level densities for
a family of nuclei using Monte Carlo sampling for a single nucleus.
In particular we can also calculate level densities of odd-even and odd-odd nuclei despite a new sign problem introduced by the projection on an 
odd number of particles. The method is applied to level densities in the
iron region using the complete $pf+g_{9/2}$-shell. The single-particle level density parameter $a$ and the
backshift parameter $\Delta$ are extracted by fitting the
microscopically calculated level densities to the backshifted
Bethe formula. We find good 
agreement with experimental level
densities with no adjustable parameters in the microscopic 
calculations. The parameter $a$ is found to vary smoothly with mass 
and does not show odd-even effects. The calculated backshift parameter 
$\Delta$ displays an odd-even staggering effect versus mass and is in
better agreement with the experimental data than are the empirical values.
\end{abstract}

\section*{Introduction}

The nucleosynthesis of
heavy elements takes place by radiative capture of neutrons ($s$ and $r$
process) and protons ($rp$ process) in competition with beta decay.
In the statistical regime, neutron and proton capture rates are
proportional to the level density of the corresponding compound
nucleus \cite{astro}.  Most theoretical models of level densities
are based on the Fermi gas model,
e.g., the Bethe formula \cite{bethe}, which describes the
exponential increase of the many-particle level density with
both excitation energy and mass number. In the backshifted
Bethe Formula (BBF) (see, e.g. Ref. \cite{conteur}),
shell corrections and two-body correlations are taken into account
empirically by introducing a backshift $\Delta$ of the ground state energy.
The BBF offers a good description of the experimentally determined level
densities when both $a$ and $\Delta$ are fitted for each
nucleus \cite{dilg,Gr74}.
The overall systematics of $a$ and $\Delta$ were studied empirically
but it is difficult to accurately predict these parameters for a 
particular nucleus.

The interacting shell model takes into account both shell effects 
and residual interactions and constitutes an attractive framework 
for calculating accurate
level densities. However, conventional diagonalization methods are
limited by the size of the model space. Full major shell
calculations are presently restricted to nuclei with 
$A \lesssim 50$ (in the $pf$-shell) \cite{shell}.
The development of quantum shell model Monte Carlo (SMMC) methods 
\cite{lang,sign}
allows the calculation of finite and zero-temperature observables in model spaces orders of magnitude larger than those that 
can be treated by conventional diagonalization
techniques. Recently the SMMC method was
successfully adapted to the microscopic calculations of nuclear
level densities \cite{nakada}.

The applications of fermionic Monte Carlo methods are often limited
by the so-called sign problem, which causes a breakdown of the method
at low temperatures. A practical solution was developed in the 
nuclear case \cite{sign}, but the resulting extrapolation errors
were found to be too large for accurate calculations of level
densities. Instead we have constructed good-sign interactions 
that include the dominating collective components of effective 
nuclear interactions \cite{dufour} and were proven to be realistic
for the calculation of level densities.

The SMMC method is based on a representation of the many-body 
imaginary-time propagator as a functional integral over one-body 
propagators in fluctuating auxiliary fields, known as the 
Hubbard-Stratonovich transformation \cite{Hubbard}.  The many-dimensional
integration is then performed by Monte Carlo. The SMMC method is
computationally intensive. In particular, level density calculations
require computation of the thermal energy at
all temperatures. If this procedure is to be repeated for a series of
nuclei, the calculations quickly become very time-consuming. Recently we
introduced a novel particle-number reprojection method \cite{reproject}
with which we can calculate nuclear observables for a series of 
nuclei using the Monte Carlo
sampling for a single nucleus.  The weight function used in the
sampling is proportional to the partition function of a fixed even-even
or $N=Z$ nucleus. Thermal observables for neighboring nuclei are then
calculated by reprojection on different particle numbers (both
even and odd). This technique offers an economical way of calculating 
level densities for a large number of nuclei.

\section*{The Shell Model Monte Carlo Methods}

A general many-body Hamiltonian containing up to two-body
interactions can be written
in the following quadratic form:
\begin{equation}
  \label{hamiltonian}
  H = \sum _\alpha \epsilon_\alpha \hat{\rho}_\alpha
  + \frac{1}{2} \sum _ \alpha v_\alpha \hat{\rho}_\alpha^2 \;,
\end{equation}
where $\hat\rho_\alpha$ are one-body densities.
Using the Hubbard-Stratonovich transformation, the imaginary-time
many-body propagator $e^{- \beta H}$ can be represented as \begin{equation}
e^{- \beta H} = \int D[\sigma] G(\sigma) U_\sigma \;,
\end{equation}
where $G(\sigma)$ is a Gaussian weight and $U_\sigma$ is a one-body
propagator of non-interacting nucleons moving in fluctuating
time-dependent  auxiliary
fields $\sigma(\tau)$. The canonical expectation value of an
observable $\hat{O}$ at inverse temperature $\beta$ is calculated from
\begin{equation}
  \label{observ}
  \rll \hat{O} \rl _ {\cal{A}}  = \frac{\int D[\sigma] 
G(\sigma) {\rm Tr}_{\cal{A}}
  (\hat{O}U_\sigma)} { \int D[\sigma] G(\sigma) {\rm Tr}_{\cal{A}}U_\sigma },
\end{equation}
where Tr$_{\cal{A}}$ is a canonical trace in the subspace of
fixed particle number ${\cal A}$. In practice we project on both
neutron and proton number, 
 $N$ and $Z$, respectively, so
$\cal{A}$ denotes the pair $(N,Z)$. Introducing the notation
  $\rll X_\sigma \rl _W \equiv \int D[\sigma] W(\sigma) X_\sigma /
   \int D[\sigma] W(\sigma)$,
where $W(\sigma) \equiv G(\sigma) {\rm Tr}_{\cal{A}} U_\sigma$, Eq.
(\ref{observ}) can be written as
\begin{equation}\label{thermal}
  \rll O \rl _{\cal A} = \rll {\rm Tr}_{\cal A}(\hat{O}U_\sigma)
  /{\rm Tr}_{\cal A}U_\sigma \rl _W.
\end{equation}
For a good-sign interaction and for
even-even or $N=Z$ nuclei, the weight function $W(\sigma)$ is
positive-definite. The $\sigma$-fields  are sampled according 
to $W(\sigma)$ and thermal observables are calculated from (\ref{thermal}).

\section*{The particle-number reprojection method}

We assume that the Monte Carlo sampling is done for a nucleus with particle
number ${\cal A}$. The ratio  $Z_{\cal A'}/Z_{\cal A}$ between
the partition function of another nucleus with particle number ${\cal A'}$
and that of the original nucleus ${\cal A}$ is written as
\begin{equation}
  \label{partitratio}
  \frac{Z_{\cal A'}(\beta)} { Z_{\cal A}(\beta)}
  \equiv
  \frac{ {\rm Tr}_{\cal A'}e^{-\beta H} } { {\rm Tr}_{\cal A} e^{-\beta H}}
  = \left \rll \frac{{\rm Tr}_{\cal A'} U_\sigma} { {\rm Tr}_{\cal A} U_\sigma}
  \right \rl _ W.
\end{equation}
The expectation value of an observable $\hat{O}$ for nucleus with
${\cal A'}$ particles is calculated from
\begin{equation}
  \label{projobserv}
  \rll \hat{O} \rl _{\cal A'} =
  \frac{ \left \rll
      \left( \frac{ {\rm Tr}_{\cal A'}\hat{O} U_\sigma} {{\rm Tr}_{\cal A'}
          U_\sigma}
      \right )
      \left( \frac{ {\rm Tr}_{\cal A'}U_\sigma} {{\rm Tr}_{\cal A} U_\sigma}
      \right ) \right \rl _W }
  { \left \rll \frac{ {\rm Tr}_{\cal A'}U_\sigma} {{\rm Tr}_{\cal A} U_\sigma}
    \right \rl _W }\;.
\end{equation}
The Monte Carlo sampling is carried out using the weight function
$W(\sigma)$ which is proportional to the partition function of 
nucleus ${\cal A}$, and
Eq. (\ref{projobserv}) is used to calculate
the same observable for nuclei with
${\cal A'} \ne {\cal A}$.

In the calculations of level densities  we used the Hamiltonian \cite{nakada}
\begin{equation}
  \label{goodhamil}
  H=\sum_a\epsilon_a\hat{n}_a + g_0 P^{(0, 1)\dagger} \cdot \tilde{P}^{(0,1)}
  - \chi \sum _{\lambda=2}^{4} k_\lambda O^{(\lambda, 0)} \cdot O^{(\lambda, 0)},
\end{equation}
where
\begin{equation}
  \begin{array}{ccl}
    P^{(\lambda, T )\dagger} & = &
    \frac{\sqrt{4\pi}}{2(2\lambda+1)}
    \sum\limits_{ab} \rll j_a \| Y_\lambda \| j_b \rl
    [ a^\dagger_{j_a} \times a^\dagger_{j_b}] ^{(\lambda, T)}, \\
    O^{(\lambda, T)} & = &
    \frac{1}{\sqrt{2\lambda+1}}
    \sum\limits_{ab} \left\rll j_a \| \frac{dV}{dr} Y_\lambda \| j_b \right \rl
    [ a^\dagger_{j_a} \times \tilde{a}^\dagger_{j_b}] ^{(\lambda, T)}.
  \end{array}
\end{equation}
The modified annihilation operator is defined by $\tilde{a}_{j, m, m_t}
= (-1)^{j-m+\frac{1}{2}-m_t} a_{j, -m, -m_t}$, and a similar definition is used for $\tilde{P}^{(\lambda, T)}$.  The single-particle energies 
$\epsilon_a$ are calculated in a central Woods-Saxon potential $V(r)$
plus spin-orbit interaction. $g_0$ is a monopole pairing
interaction strength determined from experimental odd-even mass 
differences. The quadrupole, octupole and hexadecupole interaction 
terms in (\ref{goodhamil}) are obtained by expanding a separable
surface-peaked interaction $v(r, r') = - \chi (dV/dr) (dV/dr') \delta(r-r')$
\cite{bertsch} whose strength $\chi$ is determined self-consistently.
The parameters $k_2=2,\; k_3=1.5$ and $k_4=1$ are renormalization 
constants that take into account core polarization effects.
Both the pairing and the surface-peaked interactions are attractive 
and therefore have a good sign \cite{sign}.

In the particle-number reprojection method described above we have
assumed that
the Hamiltonian $H$ is independent of $\cal{A}$.  Suitable 
corrections should be made if some of the Hamiltonian parameters 
vary with $\cal{A}$.  In the iron region we find that $\chi$
depends only weakly on the mass number $A$, and the
pairing strength $g_0$ is constant through the shell. The largest
variation in this mass region is that of the single-particle 
energies. The thermal energy  of a nucleus with ${\cal A}' =(N',Z')$ 
can then be estimated from
$E_{{\cal A}'} (\beta) \approx \sum_a [\epsilon_a({\cal A}') -\epsilon_a({\cal
A})]\langle  n_a \rangle_{{\cal A}'} + \langle H \rangle_{{\cal A}'}$,
where $H$ is the Hamiltonian for a nucleus with ${\cal A} = (N,Z)$.
This estimate  for $E_{\cal A'}$ is an approximation since we are still
using the propagator $e^{-\beta H}$ with the Hamiltonian $H$ for nucleus
${\cal A}$ (instead of ${\cal A}'$). This is a good approximation if 
we reproject on
nuclei with $N'-Z'$  values close to $N - Z$ (the Woods-Saxon potential
depends on $N-Z$).  In the applications below this leads to negligible errors in the level densities.

\section*{Applications and Results}

In this section we present applications of the particle-number 
reprojection method to nuclei in the iron region. Since we 
are interested in level densities around the neutron and proton 
resonance energies we use the complete $(pf+g_{9/2})$-shell. This model
space contains both positive and negative parity states.

We perform the direct Monte
Carlo sampling for the even-even nucleus $^{56}$Fe and the $N=Z$ nucleus 
$^{54}$Co (both have a good sign for the interaction (\ref{goodhamil})). 
The thermal
energies of $^{53-56}$Mn, $^{54-58}$Fe and
$^{54-60}$Co were reprojected from $^{56}$Fe, while those of $^{50-52}$Mn
and $^{52, 53}$Fe were reprojected from $^{54}$Co.  The calculations
were done for $\beta$ values up to
2.5 MeV$^{-1}$. At small $\beta\ ( < 1)$ the calculations were done
in a smaller step of $\Delta\beta\, =\, 1/16$. At larger $\beta$ points,
the Monte Carlo
calculations become more time-consuming, and we doubled our step size to
$\Delta\beta\, =\, 1/8$. For each $\beta$ point we took about
$4000$ independent samples. The reprojected energies usually have
larger statistical error at large values of $\beta$. To calculate 
reliable ground state energies we performed direct
Monte Carlo runs for some of the reprojected nuclei at several
values of $\beta$ between $1.75$ and $2.5$. For $\beta\ >\ 2.5$
the statistical error for the thermal energy of an odd-even nucleus 
becomes too large to be useful.
Since the thermal energy of an odd-even nucleus is already close to
its asymptotic value at these large $\beta$ values, we could extract the 
ground state energy to within an accuracy of $\sim 0.3$ MeV.

\begin{figure}\epsfxsize=9.cm
  \centerline{
    \epsffile{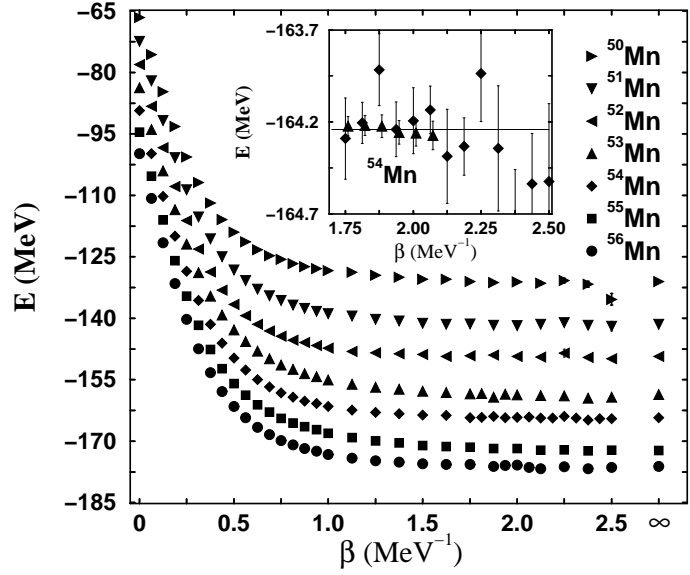}
    }
    \caption{The SMMC thermal energies vs. $\beta$ for $^{50-56}$Mn. Shown
      on the far right are the extrapolated ground-state energies.
The inset  shows the SMMC thermal
      energies (diamonds) for $^{54}$Mn at large $\beta$ values.
      The triangles are the energies obtained by averaging the
      large-$\beta$ results above the corresponding $\beta$.
      The extrapolated ground state energy is shown by a horizontal line.
      }
    \label{fig1}
\end{figure}

Fig. \ref{fig1} shows the SMMC thermal energies versus $\beta$
for manganese isotopes. The staggering observed in the spacings of the
thermal energies at large $\beta$ is a pairing effect.  The inset of
Fig. \ref{fig1} shows the SMMC thermal
energies of $^{54}$Mn (diamonds with error bars) at large values of
$\beta$. It demonstrates the procedure we used to extract the
ground state energy.

The level density is related to the partition function by an inverse Laplace
transform
\begin{equation}
  \label{laplace}
  \rho_{\cal A'}(E) = \int_{-i\infty}^{i\infty} \frac{d \beta}{2 \pi i}
  e^{\beta E_{\cal A'}} Z_{\cal A'}(\beta).
\end{equation}
The partition function $Z_{\cal A'}(\beta)$ is computed from the SMMC
thermal energies by integrating the thermodynamic relation
$-\partial \ln Z_{{\cal A}'}/\partial  \beta = E_{\cal A'}(\beta)$.
The average
level density is then calculated by evaluating (\ref{laplace})
in the saddle point method
\begin{eqnarray}\label{level}
  \rho_{{\cal A}'} = (2\pi \beta^{-2} C_{{\cal A}'})^{-1/2}  e^{S_{{\cal
        A}'}}  \;,
\end{eqnarray}
where
$S_{{\cal A}'} = \beta E_{{\cal A}'} +  \ln Z_{{\cal A}'} $
and $C_{{\cal A}'}  = - \beta^2 dE_{{\cal A}'} / d\beta$ are the 
canonical entropy and heat capacity, respectively.

\begin{figure}\epsfxsize=9.cm
  \centerline{
    \epsffile{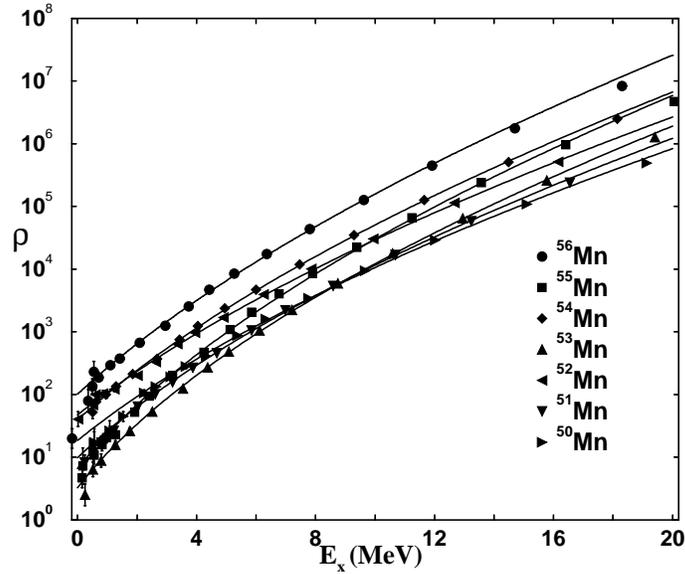}
    }
    \caption{The SMMC level densities of the $^{50-56}$Mn isotopes.
      The solid lines are the fit of the calculated level densities
      to the BBF (\protect\ref{BBF}).
      }
    \label{fig2}
\end{figure}

Fig. \ref{fig2} shows the level densities for the manganese isotopes 
of Fig. \ref{fig1} as a function of excitation energy.
These densities are fitted to a modified version of the BBF \cite{conteur}
\begin{eqnarray}\label{BBF}
  \rho (E_x) \approx {{\sqrt\pi}\over{12}} a^{-\frac{1}{4}}
  (E_x - \Delta +t)^{-\frac{5}{4}} e^{2\sqrt{a (E_x - \Delta)}} \;.
\end{eqnarray}
Here $t$ is a thermodynamic temperature determined by the relation
$E_x - \Delta = a t^2 - t$.
Eq. (\ref{BBF}) differs from the usual BBF in the term $t$ which
appears in the pre-exponential
factor, and gives a better fit to
the SMMC level densities at lower excitation energies. The solid lines
in Fig. \ref{fig2} are the fitted BBF level densities of Eq. (\ref{BBF}).
The fitting is done in the energy range $E_x < 20$ MeV and is
usually good down to $\sim 1$ MeV for even-even nuclei (for which 
$\Delta$ is positive), or even below $1$ MeV for odd-$A$ nuclei.
The reduced pairing correlations in odd-odd nuclei are clearly 
observed in the level densities of Fig. \ref{fig2}. The backshift
parameter $\Delta$ for the odd-odd nucleus $^{54}$Co is lower than
$\Delta$ for the odd-even nucleus $^{55}$Co, leading to a higher level 
density for $^{54}$Co despite its smaller mass.

\begin{figure}[h]\epsfxsize=9. cm
  \centerline{
    \epsffile{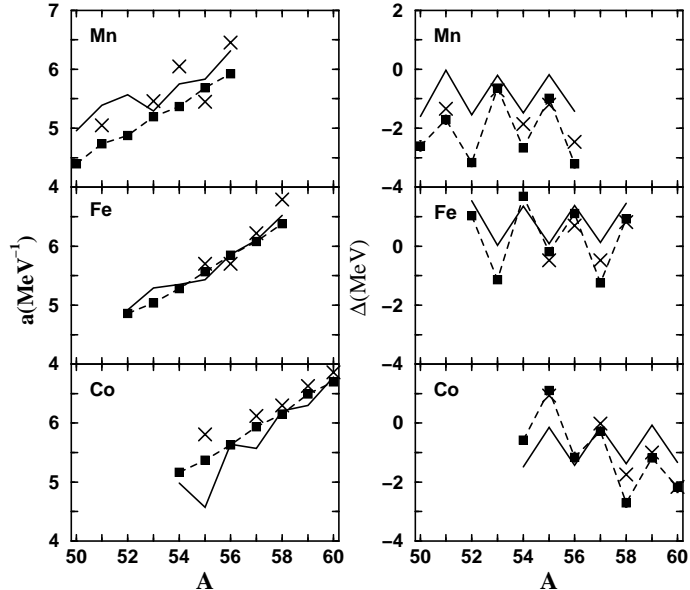}
    }
  \caption{The single-particle level density parameter $a$ (left column)
    and the backshift  parameter $\Delta$ (right column) for Mn, Fe and Co
    isotopes. The solid  squares are the SMMC values (obtained by 
fitting the SMMC level densities to the BBF
    (\protect\ref{BBF})). The $\times$'s are
    the  experimental results taken from the
    compilations of \protect\cite{dilg} (assuming rigid body moment of
    inertia), except for $^{58}$Co and $^{59}$Co where we used the
values quoted in \protect\cite{Gr74}. The solid lines are the empirical formulae
    of \protect\cite{HWFZ}. Taken from Ref. \protect\cite{reproject}.
    }
  \label{fig3}
\end{figure}

The level density parameters $a$ and $\Delta$ were extracted by 
fitting Eq. (\ref{BBF}) to the microscopic SMMC level densities, 
and are shown in Fig.
\ref{fig3} versus mass number $A$. The SMMC results (solid squares) 
are compared with
the experimental data ($\times$'s) quoted in Refs. \cite{dilg}
and \cite{Gr74}.
The solid lines describe the empirical formulae of Refs. \cite{HWFZ}.
The SMMC values of $a$ depend smoothly on the mass $A$, unlike
the values predicted by the empirical formulae.
The pairing effects are clearly reflected in the staggering behavior
of $\Delta$ versus $A$ as seen on the right column of Fig. \ref{fig3}. In 
the empirical formulae, $\Delta$ is close to zero for odd-even nuclei, 
positive for even-even
nuclei and negative for odd-odd nuclei. However, we see that both the
experimental and SMMC values of $\Delta$ can differ significantly from
zero for the odd-even nuclei. The SMMC values of $a$ and particularly
$\Delta$ are generally in better agreement with the experimental
results than the empirical values.
For some of the odd-odd manganese isotopes we observe discrepancies
between the SMMC values of $a$ and the experimental data. However, 
the lower values of $a$ for these manganese isotopes are compensated 
by corresponding
lower values of $\Delta$. Consequently, the discrepancies
in the level densities themselves are less significant for
$E_x \lesssim 10$ MeV.

\begin{figure}\epsfxsize=9.5cm
  \centerline{
    \epsffile{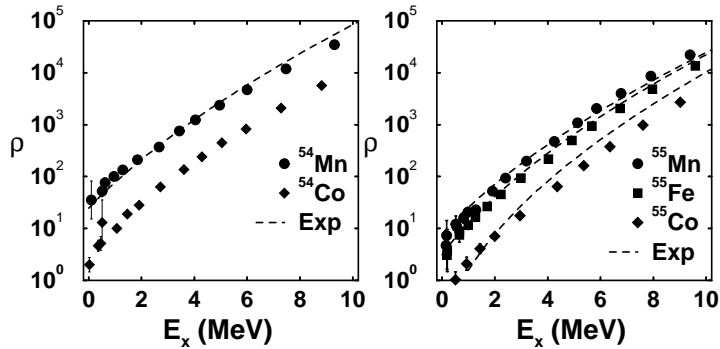}
    }
  \caption{The level densities of two odd-odd $A=54$ nuclei: 
$^{54}$Mn and $^{54}$Co (left), and three odd-even $A=55$ nuclei:
$^{55}$Mn, $^{55}$Fe and $^{57}$Co (right). The symbols are the
SMMC level densities and the dashed lines are the experimental level 
densities.
    }
  \label{fig4}
\end{figure}

To demonstrate the $T_z$-dependence of level densities we show in
Fig. \ref{fig4} the level densities of
two odd-odd $A=54$ (Mn and Co) and three odd-even $A=55$ nuclei 
(Mn, Fe and Co).
The empirical formulae
predict similar level densities for the two odd-odd nuclei as well as for the three 
odd-$A$ nuclei: the values of $a$ are similar if the
mass $A$ is the same; $\Delta \sim 0$ for odd-$A$ nuclei; and 
$\Delta$ are
approximately the same for odd-odd nuclei. However the SMMC
level densities of these
nuclei (symbols) are seen to be quite different from each other. 
We also see that the experimental level densities (dashed lines) 
are in good agreement with the SMMC densities.

In conclusion, we have described a particle-number reprojection method in the shell model Monte Carlo method. With this reprojection technique we can calculate the thermal
properties for a series of nuclei using Monte Carlo sampling for a
single nucleus.  Level densities of
odd-$A$ and odd-odd nuclei are calculated 
despite a new sign problem introduced by the projection on an odd
number of particles.

This work was supported in part by the Department of Energy grant
No.\ DE-FG-0291-ER-40608, and by the Ministry of Education, Science, Sports and  Culture of Japan (grants 08044056 and 11740137). Computational cycles
were provided by the Cornell Theory Center, by the San Diego Supercomputer Center (using NPACI resources), and
 by the NERSC high performance computing facility at LBL.

\end{document}